\documentclass[preprint,showpacs,prl]{revtex4}
\usepackage{graphics}

\begin{document}

\title{Prediction of a new class of half-metallic antiferromagnets}

\author{D. K\"odderitzsch} \affiliation{Fachbereich Physik,
  Martin-Luther-Universit\"at Halle-Wittenberg,Friedemann-Bach-Platz 6 Halle,
  Germany 06099 }
\author{W. Hergert}%
\affiliation{Fachbereich Physik, Martin-Luther-Universit\"at
  Halle-Wittenberg,Friedemann-Bach-Platz 6 Halle, Germany 06099 }
\author{Z. Szotek}%
\affiliation{Daresbury Laboratory, Daresbury, Warrington WA4 4AD, United
  Kingdom}
\author{W. M.  Temmerman}%
\affiliation{Daresbury Laboratory, Daresbury, Warrington WA4 4AD, United
  Kingdom}

\date{\today}

\begin{abstract}
  We report on vacancy induced half-metallicity in the prototype Mott-insulating
  substances MnO and NiO. By embedding a cation-vacancy into a \emph{magnetic}
  system a new road opens up to the construction of half-metallic
  antiferromagnets. For Ni$_{0.97}$O we find a half-metallic antiferromagnet a system
  hitherto only proposed for complicated crystal structures.
\end{abstract}

\maketitle

Half-metallic ferromagnets \cite{GrootMEn+83} which exhibit a gap at the Fermi
energy in one spin channel only, giving rise to 100\% spin polarisation, may
play an important role in the new field of spintronics. By virtue of the integer
filling of one spin channel in stoichiometric half-metals, the resulting total
magnetic moment (in $\mu_B$) per unit-cell is also integer. The question whether the total
moment can be zero, whilst retaining half-metallicity, has been addressed
first by Leuken \emph{et al.} \cite{LeukenG95} Not to be confused with usual
antiferromagnets they defined a half-metallic antiferromagnet (HMAF) as a
half-metal with vanishing macroscopic moment. Based on the unique properties of
such a system proposals for potential applications have been made. Half-metallic
antiferromagnets would be
ideal tip materials in spin-polarised STM (SPSTM) as the problem of operating
with a permanent magnetic tip above the surface of a sensitive magnetic system
could be circumvented. Furthermore a new application has been pointed out by
Pickett, \cite{Picket96} who argued that the problem of breakdown of
superconductivity due to high magnetic fields could be tackled by using HMAFs. In
HMAF there is no such obstacle because of the enforced vanishing spin moment.

Theoretical investigations starting from Heusler-alloys predicted for the first
time a HMAF. This was achieved by substituting Mn and In for V and Sb,
respectively, in the nonmagnetic semiconductor VFeSb. The resulting material
had, however, a complicated crystal structure. Other promising candidates
include such double perovskites as La$_{2}$VMnO$_{6}$ and
La$_{2}$VCuO$_{6}$.\cite{Picket98} In principle, HMAF materials should not be
difficult to find, however, the experimental realisation of them is still
lacking and remains a challenge.

Recently, it has been predicted \cite{ElfimovYS02} that nonmagnetic insulating
oxides, like CaO and MgO, in the rocksalt-structure, can be made half-metallic
upon introduction of a low concentration of vacancies on the cation sites.
Inspired by this and the STM evidence for cation defects in the NiO
surface, \cite{Castel+97}
in this paper we show, based on first principles calculations, that introducing
cation vacancies into antiferromagnetic insulating transition-metal (TM) oxides not
only leads to half-metallic behaviour, but in NiO a half-metallic antiferromagnetic 
state can be realised.
In fact, we speculate that a new class of HMAF materials can be engineered from 
Ni-based antiferromagnetic insulating materials.

\begin{figure}[!ht] %
{\par\centering \resizebox*{7.5cm}{4cm}{\includegraphics{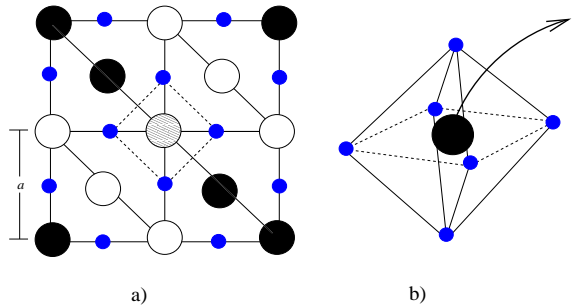}} \par}
\caption{a) Schematic view of the central layer of the supercell which corresponds to
  a (100) plane of MnO (NiO). Big  circles
  represent transition-metal sites, while additionally the spin-up and -down
  sublattices of the AF2 magnetic ordering are marked by black and empty circles,
  respectively. A cation vacancy is depicted
  in the center by a shaded circle. Small circles represent
  oxygens. b) A vacancy is created by taking out one of the transition metal
  cations in one sublattice, leaving behind a defect which is surrounded by oxygen
  in an octahedral coordination.}\label{fig1}
\end{figure}

The 3$d$ transition-metal monoxides crystalize in the rocksalt structure which
consists of two fcc sublattices displaced with respect to each other by half the
lattice constant $a$ of the cubic unit cell (see Fig.~\ref{fig1} a). Oxygen atoms
reside on one sublattice, the cations on the other. In addition, mediated by
the superexchange mechanism of Anderson-type, the transition metal ions adopt in
most of these compounds a magnetic ordering of antiferromagnetic type 2 (AF2),
where planes of spin-up and down sites are stacked in an alternating manner
along the [111] direction. The superexchange is mediated by the oxygens which by
symmetry show magnetic frustration and therefore no magnetic moment. Both the
transition metals and oxygens are situated in an octahedrally coordinated cage 
of ions of the other type (Fig.~\ref{fig1} b).
Most of these oxides are antiferromagnetic insulators even though they have
only partially filled cation- $d$-shells, due to strongly correlated nature of
the $d$-electrons in these systems.

Density functional theory (DFT) in the local-spin-density (LSD) approximation 
cannot capture the strongly correlated nature of transition metal oxides and 
hence fails in describing their correct groundstates and magnitude of energy 
gaps. \cite{Ter84a,DBS+94,OTW83} These compounds are known to be wide gap
charge transfer antiferromagnetic insulators \cite{SA84,ZSA85}, however LSD 
predicts them to be either small gap Mott-Hubbard antiferromagnetic insulators 
or even metals. On the other hand the self-interaction corrected (SIC)-LSD can 
provide an adequate description and correct groundstate of these 
materials, \cite{SG90,Szo93,Koedderitzsch+02} and it is the approach we use in 
the present work.

In the transition metal monoxides discussed here, the TM element is
divalent and contributes two $s, p$ electrons to the oxygen $p$-bands which are
then used to fill these bands.  (This corresponds in an ionic model to the nominal
configuration TM$^{2+}$O$^{2-}$, where oxygen uses the electrons to fill up its
$p$-shell). Thus, by taking a TM ion out of the antiferromagnetic
insulating material one will create two holes in the predominantly oxygen $p$ 
bands, since MnO and NiO are charge transfer insulators. \cite{SA84,ZSA85} 

To study this in a quantitative detail, we have determined the electronic structure of
MnO and NiO with low concentration of cation vacancies, in the framework of SIC-LSD. 
In particular, we have performed \emph{ab initio}
calculations using a supercell approach to simulate the effect of placing
vacancies in these compounds. The cubic supercell used in the calculations comprises 
32 formula units (Fig.~\ref{fig1}). To gain more insight, we have studied in 
addition to the experimentally observed AF2 ordering also a ferromagnetic (FM) alignment 
of spins. We have applied the SIC to the transition metal $d$-states only, and 
specifically in MnO five spin-up $d-$states and in NiO five spin-up and three
spin-down states (three t$_{2g}$ states) see a self-interaction corrected
potential, whilst all the other electrons respond to the usual LSD potential (see 
also \cite{Szo93}).

\begin{figure}[htbp]
  \resizebox*{0.49\textwidth}{6cm}{\includegraphics{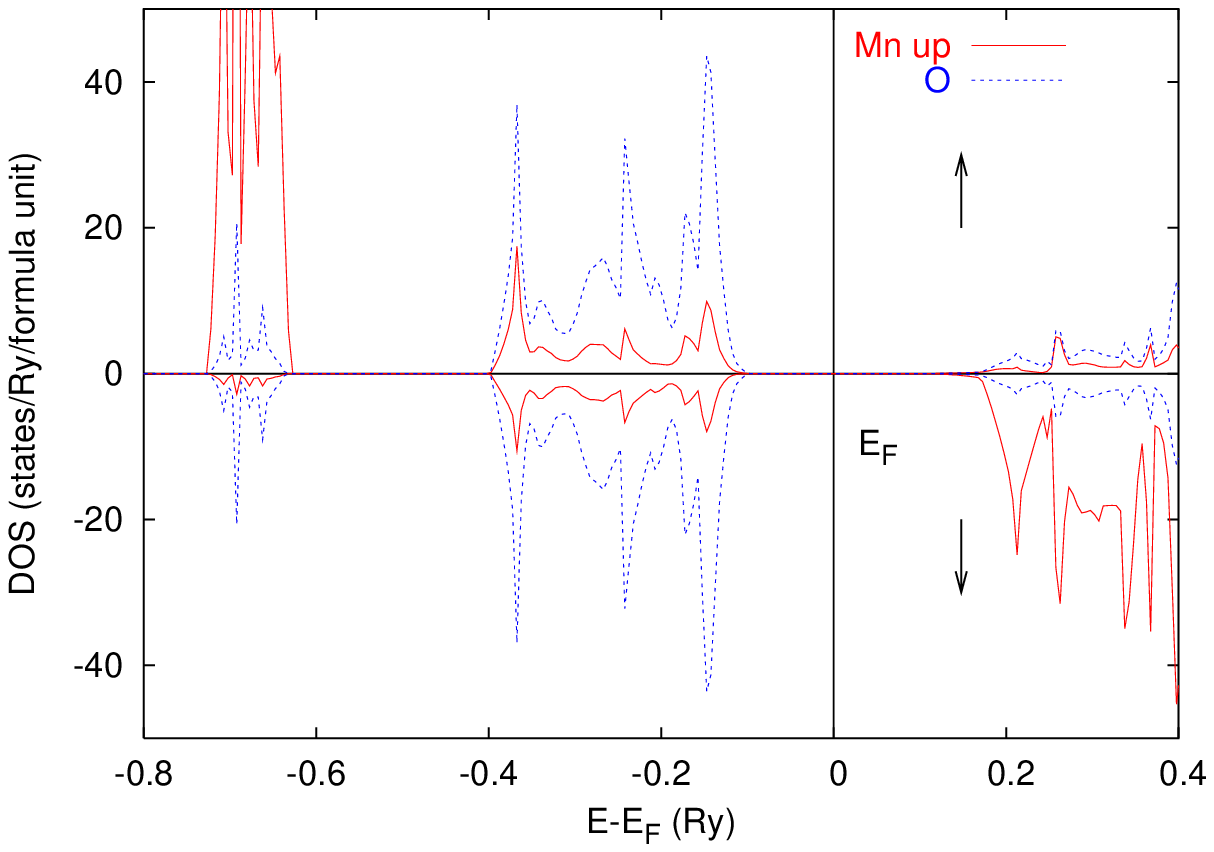}}\\
  \resizebox*{0.49\textwidth}{6cm}{\includegraphics{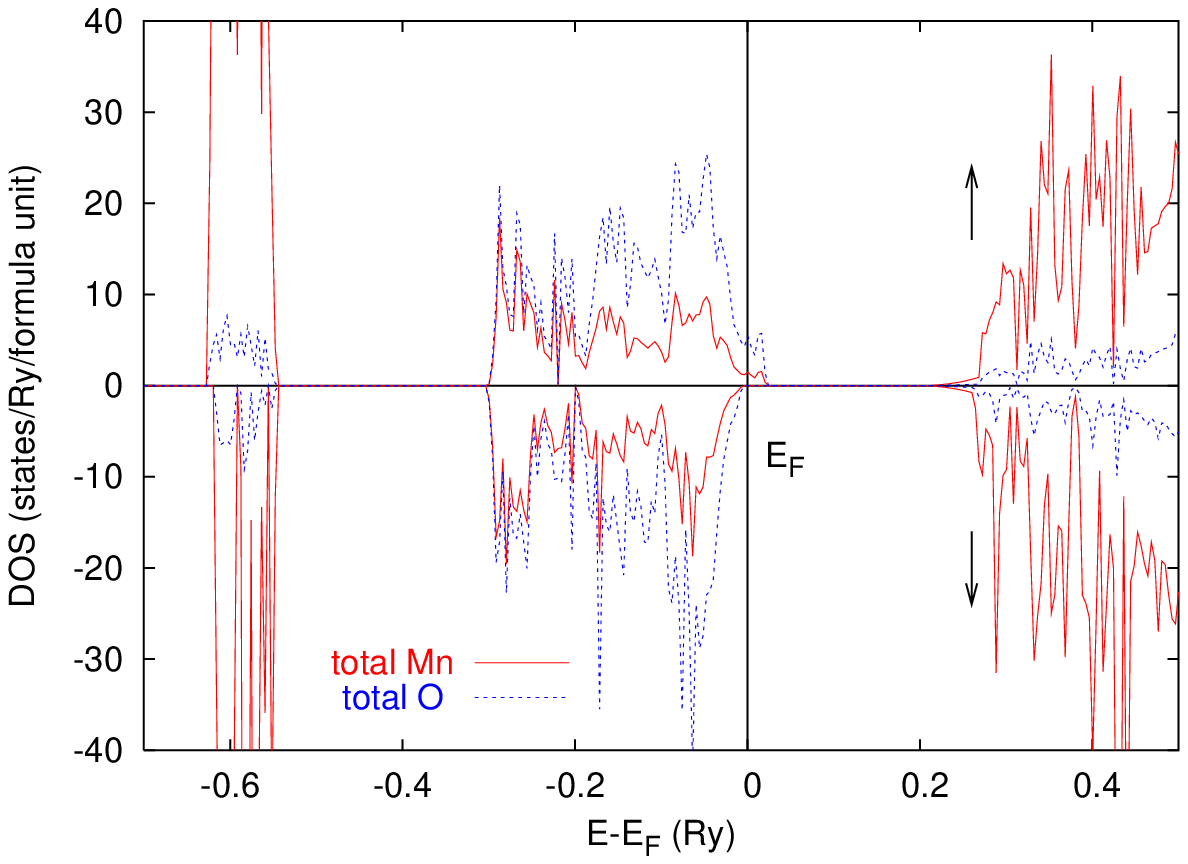}}
    \caption{Density of states (DOS) projected onto Mn-up and oxygen in bulk-MnO
      (upper panel).  Projecting onto angular momentum the oxygen DOS has most
      of its weight in the $p$-channel, whereas Mn is mainly $d$-like.  The
      partial DOS of Mn$_{0.97}$O (AF2 pattern) is shown in the lower panel.}
    \label{fig2}
\end{figure}

\begin{figure}[htbp]
  \resizebox*{0.49\textwidth}{6cm}{\includegraphics{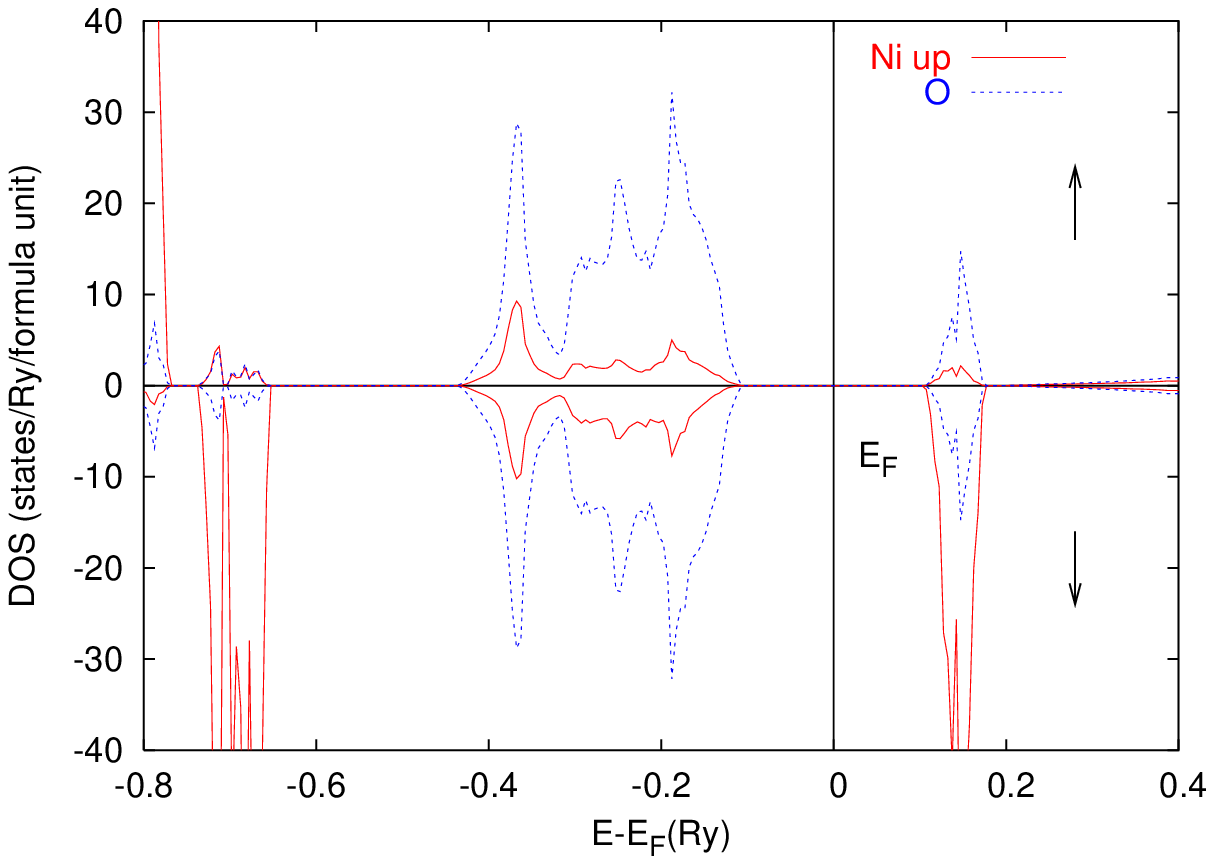}}\\
  \resizebox*{0.49\textwidth}{6cm}{\includegraphics{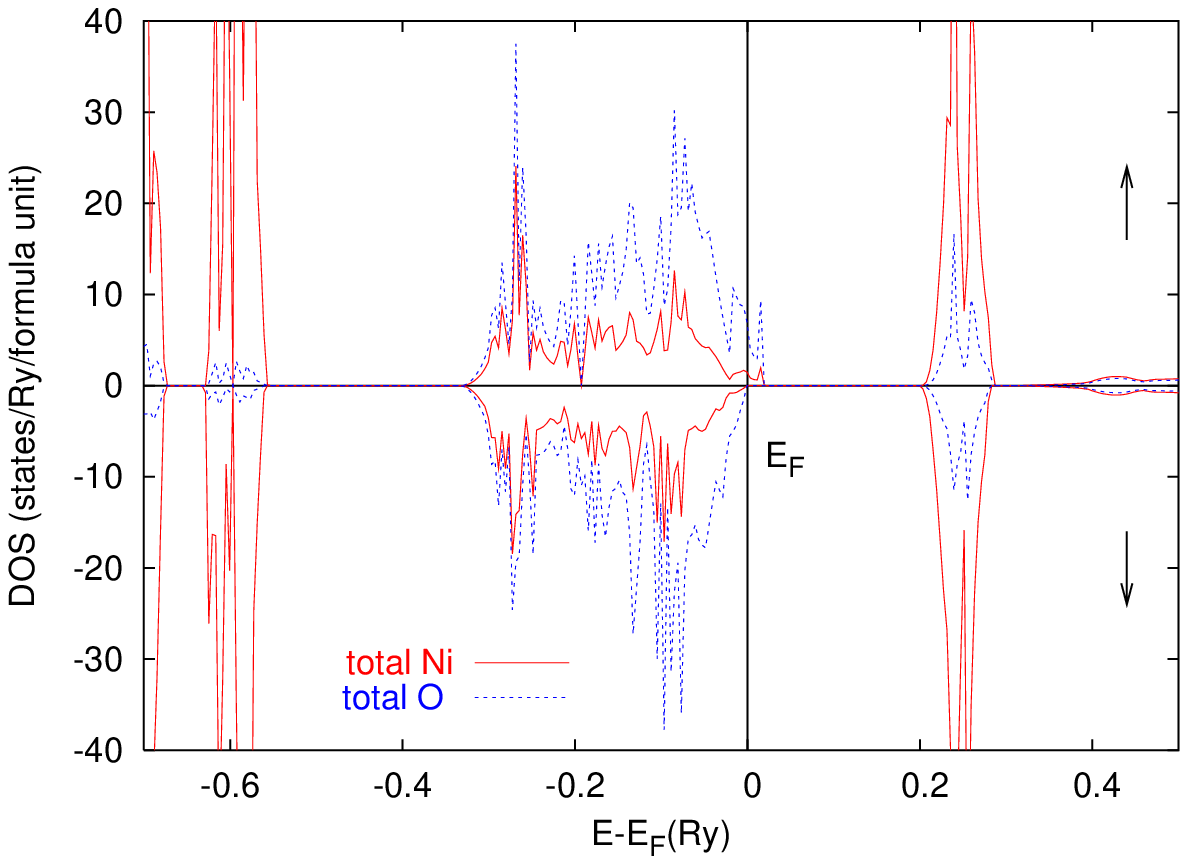}}
    \caption{DOS of bulk-NiO and Ni$_{0.97}$O (compare to Fig.~\ref{fig2}).}
    \label{fig3}
\end{figure}

In Figs.  \ref{fig2} and \ref{fig3} we show the calculated
spin-resolved densities of states (DOS) of bulk MnO and NiO, respectively, in
comparison with the respective DOS of the supercells containing vacancies, with 
the concentration of 3.125\% in the whole crystal, corresponding to Mn$_{0.97}$O 
and Ni$_{0.97}$O.
For the bulk we observe in both cases insulating behaviour with large bandgaps 
at the Fermi energy. The gaps are of charge transfer type -- the top of the valence
band is predominantly oxygen $p$-like, whereas the bottom of the conduction band
is formed by transition-metal $d$-states. The oxygen $p$ bands are not polarised
in these AF2 bulk materials.
Turning now to the case of the oxides containing vacancies we see that the Fermi
energy has moved down and crosses the valence bands. Restricting for the sake of 
argument to NiO, this can be explained as
follows. Taking out a Ni cation we remove eight 3$d$-states and 10
electrons from the compound, of which approximately eight are $d-$electrons.
The remaining two of the ten electrons are the $s, p$-electrons that have been
transferred from Ni to oxygen to fill up the $p-$band in the oxide material. 
Therefore we loose electrons in the oxygen $p$-bands which must be reflected by 
a shift of the Fermi energy into the $p-$band.
In addition, it can be seen both in Mn$_{0.97}$O and Ni$_{0.97}$O that a spin
polarisation of the predominantly oxygen $p$-like valence band has occured.
The important point here is that because of the spin-polarisation
of the oxygen $p$-bands and the chosen concentration of vacancies (i.e. the
number of lost transition metal electrons per cell) the Fermi energy crosses the
valence oxygen $p$-band in one spin channel only. Hence, both Mn$_{0.97}$O
and Ni$_{0.97}$O are half-metals, exhibiting 100\% spin-polarisation at the
Fermi-level. 
The situation is different when the concentration of vacancies is increased.
With 6.25\% vacancies on one of the magnetic sublattices in NiO, the Fermi
energy moves further down into the valence band, crossing both spin channels,
and as a consequence half-metallicity is lost.
Nevertheless, the calculated spin polarisation at the Fermi-energy is still very 
high with a value of 87\%. 
Of course, half-metallicity could also be affected by structural relaxations 
occuring as a result of introducing vacancies into these compounds. However,
our calculations for NiO with 3.125\% vacancies show that half-metallicity 
survives a local 10\% inward relaxation of the neighbouring oxygen atoms.


To gain more understanding of the magnetic properties we first discuss 
the magnetic moment formation in the ideal systems. Note that 
both in MnO and NiO the $d$-states of the free TM atoms have
nominally $S=5/2$ and $S=(5-3)/2 = 1$ configurations, respectively.  
In the monoxides 
these localised moments are reduced by hybridisation effects, such that
the moments on the Mn and Ni sites are approximately 4.5$\mu_B$ and $1.6\mu_B$, 
respectively. For ideal AF2 magnetic ordering, the oxygens are
magnetically frustrated by symmetry and have zero moment and hence, the total
moment per elementary cell is zero. If in an electronic structure calculation the
moments on the transition metal sites are forced to arrange in an FM ordering
pattern the bandgap still persists, the oxygens acquire a spin moment and the
magnetic moment per formula unit is integer, $m=2S\mu_B$, because we have filled
up bands separated by a gap (see e.g. \cite{Koedderitzsch+02})

Table \ref{tab:magn_mom_supercell_imp} shows the total magnetic moments of
MnO and NiO without and with vacancies as calculated in the present supercell approach.
We show the results for two different underlying magnetic arrangements where the
vacancy is placed into (namely AF2 and FM).
In all the cases we obtain integer magnetic
moments although for different reasons. Most importantly, note that for a vacancy 
embedded into AF2-ordered NiO, we get a total magnetic moment of zero, \emph{i.e.,}  
this system is a half-metallic antiferromagnet.

To understand how this comes about we start with the case of MnO.
Taking out one Mn cation from AF2 ordered MnO supercell would leave a surplus
magnetic moment of $5\mu_B$. However, by inspection of 
Table \ref{tab:magn_mom_supercell_imp}, this is not what is observed.
In the same way in a FM ordered system we would expect a total moment of
$155\mu_B$ and yet we get only $153\mu_B$. This implies that the spin
polarization of neighbouring atoms, induced by vacancy creation, gives
rise to a compensating integer moment of $2\mu_B$ in both the FM and AF2 cases.
For NiO with vacancy, in the AF2 ordering, due to this compensating magnetic 
moment the total magnetic moment in the unit cell is exactly zero $\mu_B$.

\begin{figure}[htbp]
  \resizebox*{0.49\textwidth}{4.5cm}{\includegraphics{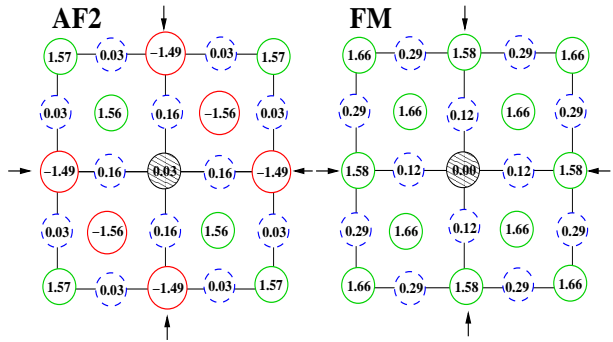}}\\
    \caption{Magnetic moments on the atomic spheres in the (100)-plane of
      NiO containing a vacancy. The underlying pattern of the supercell is AF2 (left 
panel) and
      FM (right panel). The respective moments of the ideal bulk structure are
      $m_{\mbox{\scriptsize Ni}}^{\mbox{\scriptsize AF2}}=\pm 1.56\mu_B$,
      $m_{\mbox{\scriptsize O}}^{\mbox{\scriptsize AF2}}=0\mu_B$,
      $m_{\mbox{\scriptsize Ni}}^{\mbox{\scriptsize FM}}=1.66\mu_B$, $m_{\mbox{\scriptsize
          O}}^{\mbox{\scriptsize FM}}=0.34\mu_B$ in our calcuations. Small
      arrows indicate the positions of next nearest neighbours to the impurity.}
    \label{fig4}
\end{figure}

A more detailed look at the distribution of moments around the
vacancy for the case of NiO (the findings for MnO are similar) is displayed
in Fig.~\ref{fig4} (left panel), where we show the magnetic moments in the
ASA spheres in the (100) plane in the neighbourhood of the vacancy. As can be seen
for the case of an underlying AF2 ordering the oxygens around the vacancy are
not frustrated anymore and acquire a moment of $0.16\mu_B$. In doing so they
preserve the AF2 pattern which has been destroyed by taking out the Ni atom at the
vacancy site.
When comparing the moments on the Ni spheres to the respective bulk values of $\pm
1.56\mu_B$, one realises that the nearest Ni-neighbours to the vacancy are
essentially unaffected. However, the moments on the Ni next-nearest neighbours
are decreased in magnitude by $0.07\mu_B$. This is a result of the fact that the
Anderson exchange via oxygen, which we miss out at the site of the vacancy,
plays a dominant role over the ninety degree exchange to nearest
neighbours.  This is also reflected through the change of other quantities, e.g.
the change of calculated charges in the ASA spheres. Adding up the magnetic
moments of the spheres over the whole unit cell gives zero total magnetic moment. The
compensating moment of $2\mu_B$ for the Ni-cation taken out is mainly
distributed over the oxygens surrounding the vacancy and the Ni atoms being 
the next-nearest
neighbours to the vacancy (marked by arrows in Fig.~\ref{fig4}).

The same can be observed in the case of an underlying FM ordering
(Fig.~\ref{fig4}, right panel). Here, due to the FM ordering the oxygens
are polarised already in the bulk. The compensation of $2\mu_B$ manifests itself
in the smaller magnitude of the moments on the oxygen spheres surrounding the
vacancy and the next-nearest neighbour cations.

In conclusion, we have shown vacancy induced half-metallicity in
antiferromagnetic TM oxides with the rocksalt structure. We argue, that by taking
out a cation from the lattice the induced spin-polarisation in the oxygen
$p$-band compensates an integer number of $-2\mu_B$ for the ion taken out, with
moment of 2$S\mu_B $.  In particular, we find for NiO ($S$=1), by virtue of cancellation,
a half-metallic antiferromagnet. As the observed mechanism appears to be fairly
general, we would expect it to work in other materials exhibiting an antiferromagnetic
insulating behaviour. A possible candidate might be La$_{2}$NiO$_{4}$ which is
an antiferromagnetic insulator. \cite{GT89} In the NiO$_{2}$ planes of this
compound, the oxygen atoms carry no magnetic moments, but the apex oxygen can
become slightly spin-polarized. Therefore, we would expect that a similar 
scenario could be operational in this compound. The insulating NiI$_{2}$ \cite{DBS95}
and other Ni halides could also be possible candidates for antiferromagnetic
half-metals. The conditions have to be such that the induced exchange splitting 
of the full anion band is sufficiently large to ensure that the holes are fully
spin-polarized and the spin-polarization fully compensates for the 
magnetization induced by the creation of the transition metal vacancies.
Only Ni compounds have this property, since the two holes created by a
Ni vacancy can compensate the 2 $\mu_{B}$ magnetization change of the 
vacancy. Therefore, the search for half-metallic antiferromagnets should
span all Ni oxide-based antiferromagnetic insulators as well as the more
complicated Ni halides. 
\acknowledgments{
This work was supported by the DFG through the Forschergruppe ``Oxidic
Interfaces''
and has been partially funded by the RTN "Computational
Magnetoelectronics" (HPRN-CT-2000-00143) and the ESF Psi-k Programme (STRUC).}



\begin{thebibliography}{16}
\expandafter\ifx\csname natexlab\endcsname\relax\def\natexlab#1{#1}\fi
\expandafter\ifx\csname bibnamefont\endcsname\relax
  \def\bibnamefont#1{#1}\fi
\expandafter\ifx\csname bibfnamefont\endcsname\relax
  \def\bibfnamefont#1{#1}\fi
\expandafter\ifx\csname citenamefont\endcsname\relax
  \def\citenamefont#1{#1}\fi
\expandafter\ifx\csname url\endcsname\relax
  \def\url#1{\texttt{#1}}\fi
\expandafter\ifx\csname urlprefix\endcsname\relax\def\urlprefix{URL }\fi
\providecommand{\bibinfo}[2]{#2}
\providecommand{\eprint}[2][]{\url{#2}}

\bibitem[{\citenamefont{de~Groot et~al.}(1983)\citenamefont{de~Groot, Mueller,
  van Engen, and Buschow}}]{GrootMEn+83}
\bibinfo{author}{\bibfnamefont{R.~A.} \bibnamefont{de~Groot}},
  \bibinfo{author}{\bibfnamefont{F.~M.} \bibnamefont{Mueller}},
  \bibinfo{author}{\bibfnamefont{P.~G.} \bibnamefont{van Engen}},
  \bibnamefont{and} \bibinfo{author}{\bibfnamefont{K.~H.~J.}
  \bibnamefont{Buschow}}, \bibinfo{journal}{Phys. Rev. Lett.}
  \textbf{\bibinfo{volume}{50}}, \bibinfo{pages}{2024} (\bibinfo{year}{1983}).

\bibitem[{\citenamefont{van Leuken and de~Groot}(1995)}]{LeukenG95}
\bibinfo{author}{\bibfnamefont{H.}~\bibnamefont{van Leuken}} \bibnamefont{and}
  \bibinfo{author}{\bibfnamefont{R.~A.} \bibnamefont{de~Groot}},
  \bibinfo{journal}{Phys. Rev. Lett.} \textbf{\bibinfo{volume}{74}},
  \bibinfo{pages}{1171} (\bibinfo{year}{1995}).

\bibitem[{\citenamefont{Pickett}(1996)}]{Picket96}
\bibinfo{author}{\bibfnamefont{W.~E.} \bibnamefont{Pickett}},
  \bibinfo{journal}{Phys. Rev. Lett.} \textbf{\bibinfo{volume}{77}},
  \bibinfo{pages}{3185} (\bibinfo{year}{1996}).

\bibitem[{\citenamefont{Pickett}(1998)}]{Picket98}
\bibinfo{author}{\bibfnamefont{W.~E.} \bibnamefont{Pickett}},
  \bibinfo{journal}{Phys. Rev. B} \textbf{\bibinfo{volume}{57}},
  \bibinfo{pages}{10613} (\bibinfo{year}{1998}).

\bibitem[{\citenamefont{Elfimov et~al.}(2002)\citenamefont{Elfimov, Yunoki, and
  Sawatzky}}]{ElfimovYS02}
\bibinfo{author}{\bibfnamefont{I.~S.} \bibnamefont{Elfimov}},
  \bibinfo{author}{\bibfnamefont{S.}~\bibnamefont{Yunoki}}, \bibnamefont{and}
  \bibinfo{author}{\bibfnamefont{G.~A.} \bibnamefont{Sawatzky}},
  \bibinfo{journal}{Phys. Rev. Lett.} \textbf{\bibinfo{volume}{89}},
  \bibinfo{pages}{216403} (\bibinfo{year}{2002}).

\bibitem[{\citenamefont{Castell et~al.}(1997)\citenamefont{Castell, Wincott,
  Condon, Muggelberg, Thornton, Dudarev, Sutton, and Briggs}}]{Castel+97}
\bibinfo{author}{\bibfnamefont{M.~R.} \bibnamefont{Castell}},
  \bibinfo{author}{\bibfnamefont{P.~L.} \bibnamefont{Wincott}},
  \bibinfo{author}{\bibfnamefont{N.~G.} \bibnamefont{Condon}},
  \bibinfo{author}{\bibfnamefont{C.}~\bibnamefont{Muggelberg}},
  \bibinfo{author}{\bibfnamefont{G.}~\bibnamefont{Thornton}},
  \bibinfo{author}{\bibfnamefont{S.~L.} \bibnamefont{Dudarev}},
  \bibinfo{author}{\bibfnamefont{A.~P.} \bibnamefont{Sutton}},
  \bibnamefont{and} \bibinfo{author}{\bibfnamefont{G.~A.~D.}
  \bibnamefont{Briggs}}, \bibinfo{journal}{Phys. Rev. B}
  \textbf{\bibinfo{volume}{55}}, \bibinfo{pages}{7859} (\bibinfo{year}{1997}).

\bibitem[{\citenamefont{Terakura et~al.}(1984)\citenamefont{Terakura, Oguchi,
  Williams, and Kubler}}]{Ter84a}
\bibinfo{author}{\bibfnamefont{K.}~\bibnamefont{Terakura}}, ,
  \bibinfo{author}{\bibfnamefont{T.} \bibnamefont{Oguchi}},
  \bibinfo{author}{\bibfnamefont{A.~R.} \bibnamefont{Williams}},
  \bibnamefont{and}
  \bibinfo{author}{\bibfnamefont{J.}~\bibnamefont{Kubler}},
  \bibinfo{journal}{Phys. Rev. B} \textbf{\bibinfo{volume}{30}},
  \bibinfo{pages}{4734} (\bibinfo{year}{1984}).

\bibitem[{\citenamefont{Dufek et~al.}(1994)\citenamefont{Dufek, Blaha, Sliwko,
  and Schwarz}}]{DBS+94}
\bibinfo{author}{\bibfnamefont{P.}~\bibnamefont{Dufek}},
  \bibinfo{author}{\bibfnamefont{P.}~\bibnamefont{Blaha}},
  \bibinfo{author}{\bibfnamefont{V.}~\bibnamefont{Sliwko}}, \bibnamefont{and}
  \bibinfo{author}{\bibfnamefont{K.}~\bibnamefont{Schwarz}},
  \bibinfo{journal}{Phys. Rev. B} \textbf{\bibinfo{volume}{49}},
  \bibinfo{pages}{10170} (\bibinfo{year}{1994}).

\bibitem[{\citenamefont{Oguchi et~al.}(1983)\citenamefont{Oguchi, Terakura, and
  Williams}}]{OTW83}
\bibinfo{author}{\bibfnamefont{T.}~\bibnamefont{Oguchi}},
  \bibinfo{author}{\bibfnamefont{K.}~\bibnamefont{Terakura}}, \bibnamefont{and}
  \bibinfo{author}{\bibfnamefont{A.~R.}~\bibnamefont{Williams}},
  \bibinfo{journal}{Phys. Rev. B} \textbf{\bibinfo{volume}{28}},
  \bibinfo{pages}{6443} (\bibinfo{year}{1983}).

\bibitem[{\citenamefont{Sawatzky and Allen}(1984)}]{SA84}
\bibinfo{author}{\bibfnamefont{G.~A.}~\bibnamefont{Sawatzky}} \bibnamefont{and}
  \bibinfo{author}{\bibfnamefont{J.~W.}~\bibnamefont{Allen}},
  \bibinfo{journal}{Phys. Rev. Lett.} \textbf{\bibinfo{volume}{53}},
  \bibinfo{pages}{2339} (\bibinfo{year}{1984}).

\bibitem[{\citenamefont{Zaanen et~al.}(1985)\citenamefont{Zaanen, Sawatzky, and
  Allen}}]{ZSA85}
\bibinfo{author}{\bibfnamefont{J.}~\bibnamefont{Zaanen}},
  \bibinfo{author}{\bibfnamefont{G.~A.}~\bibnamefont{Sawatzky}}, \bibnamefont{and}
  \bibinfo{author}{\bibfnamefont{J.~W.}~\bibnamefont{Allen}},
  \bibinfo{journal}{Phys. Rev. Lett.} \textbf{\bibinfo{volume}{55}},
  \bibinfo{pages}{418} (\bibinfo{year}{1985}).

\bibitem[{\citenamefont{Svane and Gunnarsson}(1990)}]{SG90}
\bibinfo{author}{\bibfnamefont{A.}~\bibnamefont{Svane}} \bibnamefont{and}
  \bibinfo{author}{\bibfnamefont{O.}~\bibnamefont{Gunnarsson}},
  \bibinfo{journal}{Phys. Rev. Lett.} \textbf{\bibinfo{volume}{65}},
  \bibinfo{pages}{1148} (\bibinfo{year}{1990}).

\bibitem[{\citenamefont{Szotek et~al.}(1993)\citenamefont{Szotek,
  Temmerman, and Winter}}]{Szo93}
\bibinfo{author}{\bibfnamefont{Z.}~\bibnamefont{Szotek}},
  \bibinfo{author}{\bibfnamefont{W.~M.}~\bibnamefont{Temmerman}}, \bibnamefont{and}
  \bibinfo{author}{\bibfnamefont{H.}~\bibnamefont{Winter}},
  \bibinfo{journal}{Phys. Rev. B} \textbf{\bibinfo{volume}{47}},
  \bibinfo{pages}{4029} (\bibinfo{year}{1993}).

\bibitem[{\citenamefont{K\"odderitzsch
  et~al.}(2002)\citenamefont{K\"odderitzsch, Hergert, Temmerman, Szotek, Ernst,
  and Winter}}]{Koedderitzsch+02}
\bibinfo{author}{\bibfnamefont{D.}~\bibnamefont{K\"odderitzsch}},
  \bibinfo{author}{\bibfnamefont{W.}~\bibnamefont{Hergert}},
  \bibinfo{author}{\bibfnamefont{W.~M.} \bibnamefont{Temmerman}},
  \bibinfo{author}{\bibfnamefont{Z.}~\bibnamefont{Szotek}},
  \bibinfo{author}{\bibfnamefont{A.}~\bibnamefont{Ernst}}, \bibnamefont{and}
  \bibinfo{author}{\bibfnamefont{H.}~\bibnamefont{Winter}},
  \bibinfo{journal}{Phys. Rev. B} \textbf{\bibinfo{volume}{66}},
  \bibinfo{pages}{64434} (\bibinfo{year}{2002}).

\bibitem[{\citenamefont{Guo et~al.}(1989)\citenamefont{Guo, , and
  Temmerman}}]{GT89}
\bibinfo{author}{\bibfnamefont{G.~Y.} \bibnamefont{Guo}} \bibnamefont{and}
  \bibinfo{author}{\bibfnamefont{W.~M.} \bibnamefont{Temmerman}},
  \bibinfo{journal}{Phys. Rev. B} \textbf{\bibinfo{volume}{40}},
  \bibinfo{pages}{285} (\bibinfo{year}{1989}).

\bibitem[{\citenamefont{Dufek et~al.}(1995)\citenamefont{Dufek, Blaha, and
  Schwarz}}]{DBS95}
\bibinfo{author}{\bibfnamefont{P.}~\bibnamefont{Dufek}},
  \bibinfo{author}{\bibfnamefont{P.}~\bibnamefont{Blaha}}, \bibnamefont{and}
  \bibinfo{author}{\bibfnamefont{K.}~\bibnamefont{Schwarz}},
  \bibinfo{journal}{Phys. Rev. B} \textbf{\bibinfo{volume}{51}},
  \bibinfo{pages}{4122} (\bibinfo{year}{1995}).

\end{thebibliography}

\newpage
\begin{table}[htbp]
  \caption{Total magnetic moments of the supercell with and without the vacancy using
    different underlying magnetic patterns (FM - ferromagnetic, AF2 -
    antiferromagnetic ordering of type 2).
    IS and HM denote insulating and half-metallic states, respectively. The formation  of the
    total magnetic moment as it is formed in the supercell derived for the ideal
    bulk is given in parenthesis.}
  \centering
  \begin{tabular}{lllllc}\hline\hline
        &            & \multicolumn{3}{c}{$m^{total}$ in $\mu_B$}& state\\\hline
    MnO & AF2 \emph{bulk}   &  & 0                &                      &IS\\
        & AF2 with vacancy  &  & 3                &                      &HM\\
        & FM \emph{bulk}    &  & 160$(=32\cdot$5) &                      &IS\\
        & FM with vacancy   &  & 153              &                      &HM\\
    NiO & AF2 \emph{bulk}   &  & 0                &                      &IS\\
        & AF2  with vacancy &  & 0                &                      &HM\\
        & FM \emph{bulk}    &  & 64$(=32\cdot$2)  &                      &IS\\
        & FM  with vacancy  &  & 60               &                      &HM\\
    \hline\hline
  \end{tabular}
  \label{tab:magn_mom_supercell_imp}
\end{table}

\end{document}